\documentclass[twocolumn]{aastex62} 

\usepackage{amsmath} 
\usepackage[whole]{bxcjkjatype} 

\newcommand{\co}[2]{CO $J=$ #1 -- #2}
\newcommand{\sio}[2]{SiO $J=$ #1 -- #2}

\newcommand{\arcdegsp}{\arcdeg\,}
\begin{document}

\title{Ejection History of the IRAS 04166$+$2706 Molecular Jet}
\author{Liang-Yao Wang（王亮堯）}
\affiliation{Academia Sinica, Theoretical Institute for Advanced Research in Astrophysics}
\affiliation{Academia Sinica, Institute of Astronomy and Astrophysics, Taipei, Taiwan}
\author{Hsien Shang（尚賢）}
\affiliation{Academia Sinica, Theoretical Institute for Advanced Research in Astrophysics}
\affiliation{Academia Sinica, Institute of Astronomy and Astrophysics, Taipei, Taiwan}
\author{Tzu-Yang Chiang（江子揚）}
\affiliation{Academia Sinica, Theoretical Institute for Advanced Research in Astrophysics}
\affiliation{Academia Sinica, Institute of Astronomy and Astrophysics, Taipei, Taiwan}
\email{shang@asiaa.sinica.edu.tw}

\date{\today}

\begin{abstract}
	
The high-velocity molecular jet driven by Class 0 protostar IRAS 04166$+$2706 exhibits a unique saw-tooth velocity pattern. It consists of a series of well-aligned symmetric knots with similar averaged speeds,  whose speeds at peaks of emission decreases roughly linearly away from the origin. Recent ALMA observations of knots R6 and B6 reveal kinematic behavior with expansion velocity increasing linearly from the axis to the edge. This pattern can be formed by a spherically expanding wind with axial density concentration. In this picture, the diverging velocity profile naturally possesses an increasing expansion velocity away from the axis, resulting in a tooth-like feature on the position--velocity diagram through projection. Such geometric picture predicts a correspondence between the slopes of the teeth and the outflow inclination angles, and the same inclination angle of $52\arcdeg$ of the IRAS 04166$+$2706 can generally explain the whole pattern. Aided by numerical simulations in the framework of unified wind model by Shang et al. (2006), the observed velocity pattern can indeed be generated. A proper geometrical distribution of the jet and wind material is essential to the reconstruction the ejection history of the system.

\end{abstract}


\section{Introduction}

Mass outflow phenomena have been an integral part of the star formation picture since the first discovery of Herbig-Haro objects in 1950s \citep{herbig1951,haro1952}.  During the gravitational collapse of dense molecular cloud, a rotating disk forms around the protostar along with the infalling gas. Theoretical considerations suggest that both mass accretion onto the protostar and the driving of a high-velocity jet from the system may occur through the inner disk under the influence of magnetic fields \citep[e.g.,][]{shu2000,bally2016}. Studies of outflows may therefore help shed light on the accretion activities of a young protostellar system.

Observations have revealed increasingly more details of outflow morphology and kinematics. They have been generally classified as parabolic shell-like or collimated jet-like according to their morphology, which also motivated two types of theoretical models. The group of outflows associated with the youngest Class 0 protostars \citep{andre1993} appear to be more collimated in general and possessing additional emission peaks at higher velocities than the classical outflows featuring broad wings in the spectra. Such component is referred to as the extremely high velocity (EHV) component to be distinguished from the classical, standard high-velocity gas (SHV) \citep{bachiller1996}. High resolution interferometric mapping further revealed that the EHV gas forms highly collimated jet-like structures along the axis while the SHV gas forms wider, cavity-like structures surrounding the former.  Examples of such Class 0 outflows showing both EHV and SHV components include L1448C \citep[e.g.,][]{hirano2010}, HH 211 \citep[e.g.,][]{gueth1999}, and HH 212 \citep[e.g.,][]{lee2007}. The simultaneous presence of a high-velocity jet and low-velocity shell in the young outflows motivate models that can explain the formation of both features like the unified wind model of \citet{shang2006}.  
	
A knotty appearance is common to the EHV jet in the youngest Class 0 outflows, which are resolved into a series of well-aligned emission peaks under high resolution observations through the SMA \citep{ho2004} or ALMA. One of the plausible knot-forming mechanism is through a variable velocity history of the flow. Signatures of shocks have been detected in some sources, which suggests an initial variability in the ejections. Some proposed physical origins for the quasi-periodic variability include: stellar magnetic cycles or global magnetospheric relaxations of the star-disk system (of few to tens of years time scale), potential perturbations by binary companions, or EXOr-FUOr outbursts \citep[see][and the references therein]{frank2014}.
	
The molecular outflow driven by the class 0 protostar IRAS 04166$+$2706 (hereafter I04166) was the first highly collimated, extremely high velocity bipolar outflow identified in the nearby Taurus molecular cloud (140 pc) \citep{tafalla2004}. Its \co{2}{1} emission show an overall length-to-width ratio of $>10$, and its EHV peaks are clearly detected in the spectra at velocities of $>30$ km\,s$^{-1}$, higher than the classical outflow wings of $<10$ km\,s$^{-1}$ relative to the system velocity. The two spectral features are loosely connected by rather weak intermediate emissions. 

The rather distinct morphology of the two components was revealed by the observations of \citet{santiago-garcia2009}. At low velocities, a symmetric pair of red and blue-shifted V-shaped cones are identified with their apex overlapping near the position of the central driving source. On the other hand, a series of well aligned emission peaks are found in the high-velocity channels, forming a collimated jet-like structure. Even considering the gradually increasing size of the high-velocity peaks, its $\simeq10$\arcdegsp widening is still much smaller than the $\simeq32$\arcdegsp opening angle of the low-velocity cones. Because of its relatively clean appearance, I04166 may be regarded as an archetypical example of the youngest molecular outflows. 
	
\begin{figure}    
\includegraphics[width=0.47\textwidth]{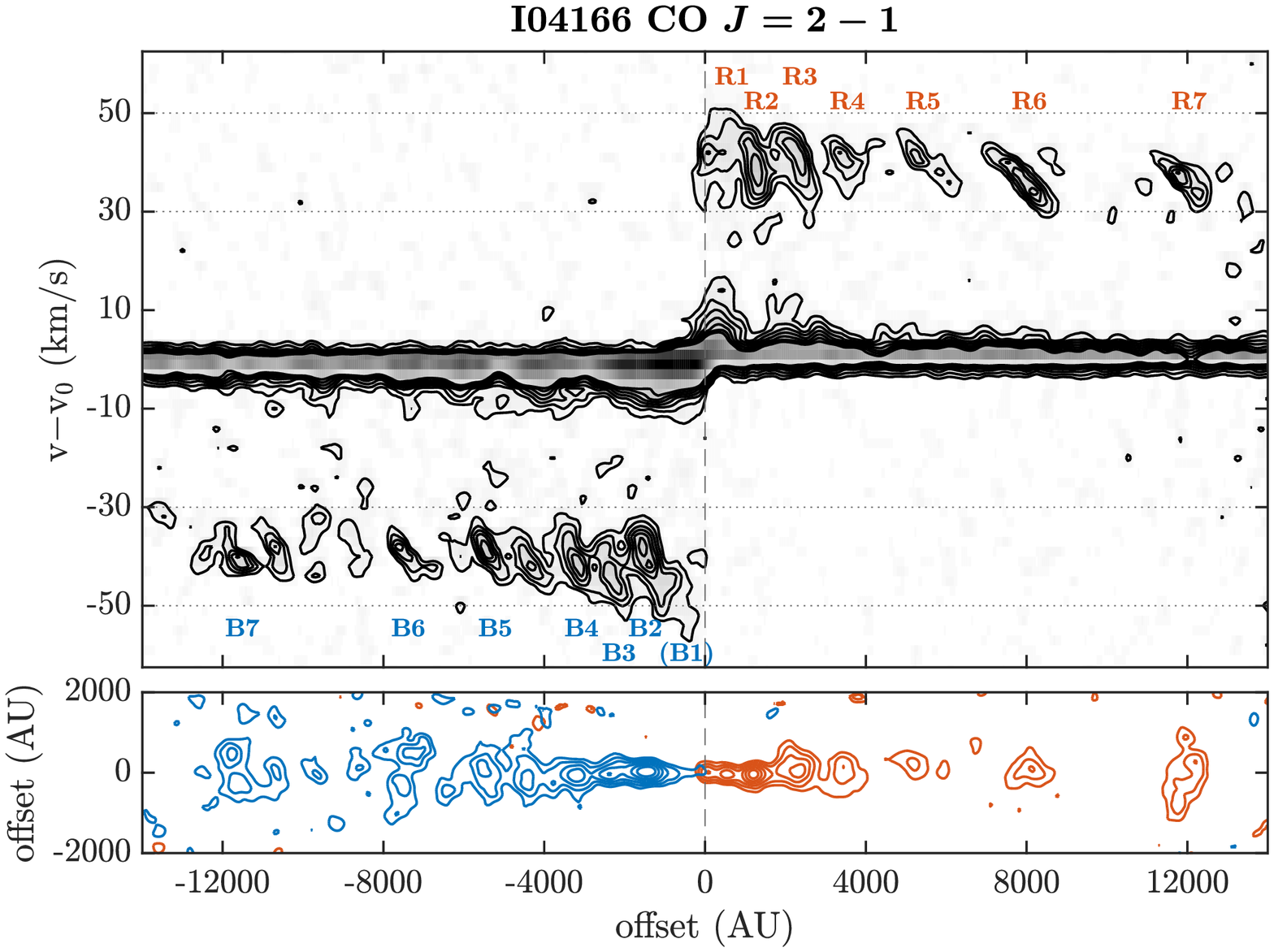}
\caption{\co{2}{1} PV diagram (\textit{upper}) and EHV emission maps (\textit{lower}) of the I04166 molecular outflow replotted from data of \citet{santiago-garcia2009}. We adopt the same labels for knots.  In the position--velocity diagram cut along the outflow axis, a sawtooth-like velocity pattern is revealed in the EHV regime of $50>|v-v_0|>30$ km\,s$^{-1}$. The first contour and level step are $0.15$ and $0.1$ Jy beam$^{-1}$. The red and blue contours in the lower panel show emissions integrated over the red and blue-shifted EHV channels, respectively. The first contour and level step are $2$ and $1$ Jy beam$^{-1}$ km s$^{-1}$.}
\label{fig:i04166_obs}
\end{figure}

More interestingly, a peculiar kinematic pattern is revealed in the position--velocity (PV) diagram cut along the axis of the I04166 outflow \citep{santiago-garcia2009}. We plot the same PV diagram in Figure \ref{fig:i04166_obs}, which clearly shows that the EHV emission consists of a series of peaks like a sawtooth. While having a similar average speed of $\sim40$ km\,s$^{-1}$, each peak possesses an internal gradient where the observed velocity drops about linearly with distance from the driving source. The ``slow head''--``fast tail'' behavior found in each knot entity is not easily understood since one would naively expect to see higher velocity gas near the head simply because it is faster.   

The discrete emission peaks of the I04166 jet suggests variable or episodic ejection activities in the past. Inferring the velocity history of the system is, however, not trivial. This is because an interpretation of the observed line-of-sight velocity will depend on a correct understanding of the flow geometry. Clues of the I04166 jet geometry may come from the two extended knots B6 and R6, whose kinematics have been mapped in details using the Atacama Large Millimeter/submillimeter Array (ALMA) by \citet{tafalla2017}. Their modeling showed that the observed velocity field is consistent with a slightly parabolic disk with expansion velocity increasing linearly from the center to the edge. They are consistent with the same interpretation of \citet{wilson1984}, \citet{raga1990b}, and \citet{stone1993b}, in which the expansion of a knot is driven by high pressure, shocked gas in an internal working surface. Numerical simulations of pulsed jets can produce a sawtooth-like velocity pattern similar to that observed in the I04166 outflow \citep[see Figure 16 of][]{stone1993b}, with a higher jet velocity (235 km\,s$^{-1}$) and high gas temperature ($10^4$\,K).
	
In this work, we propose an alternative framework of a spherical wind with mass concentration around the outflow axis. Not only can it naturally explain the tooth-like velocity pattern of individual knots, the systemic variation of tooth slopes with distance across the entire jet is also automatically reproduced. Although the notion that a molecular jet presents a spherical wind-like velocity field may seem contradicting, this is actually the expected property in magnetocentrifugal wind theories like the X-wind model \citep{shu2000}. Besides formulating an analytic description of the sawtooth velocity pattern, we also carry out numerical simulations in framework of the unified wind model of \citet{shang2006}. The model consists of a purely radially directed wind with cylindrically stratified density profile that is highly concentrated toward the axis. This is the asymptotic behavior predicted by the X-wind model and turns out to be ideal for explaining the I04166 jet.

Several observations toward the I04166 outflow are critical leading to current work. \citet{santiago-garcia2009} observed the I04166 outflow with the IRAM Plateau de Bure interferometer (PdBI) in its CD configuration during 2004--05 (blue lobe) and 2005--06 (red lobe). \co{2}{1} (230.5 GHz) and \sio{2}{1} (86.8 GHz) emissions were mapped at angular resolutions of between 2\arcsec\, and 4\arcsec. Single-dish data were added to the visibility to improve the quality of the \co{2}{1} images, and the beam-size of the combined data is $3\arcsec$. 
\citet{tafalla2017} observed the two knots B6 and R6 of the I04166 outflow at 230.5 GHz using ALMA in December 2014. The array was in its C$32-2$ configuration and no observations with the compact Morita Array were made because of the low declination of the source. \co{2}{1}, \sio{5}{4} and SO $J_N =$ $6_5$--$5_4$ transitions were observed simultaneously, and the beam-size of the CO data is $1\farcs5\times1\farcs1$. The \co{3}{2} data of \citet{wang2014} taken in 2010 by the SMA covered the inner four pairs of knots and the synthesized beam-size is $\sim1\farcs0\times0\farcs8$.

We organize the paper as follows. In Section \ref{sec:model}, we explain how the proposed model explains the observed velocity pattern. An analytic description of the pattern slope is developed, which relates the slope of the tooth pattern to the underlying outflow inclination angle. In Section \ref{sec:numerical_simulation}, we present numerical simulations based on the framework of unified wind model and realistically reproduce the observed pattern. Implications on the outflow velocity history are discussed in Section \ref{sec:discussion}. Section \ref{sec:conclusion} summarizes the results.

\section{Origin of the Velocity Pattern: An Analytic View}

\label{sec:model}

The nature of the sawtooth velocity pattern on the Position-Velocity diagram (PV Diagram) observed in the I04166 jet has posted an intriguing challenge to models. Each emission peaks of both the blue and red shifted lobes presents an internal velocity gradient of ``slow head--fast tail'' appearance on the PV diagram along the outflow axis. Moreover, the slopes of these tooth features (the apparent velocity gradient on the PV diagram) appear to vary systematically with distance from the driving source, where the inner ones are steeper than the outer ones (Figure \ref{fig:i04166_obs}). Such ordered behavior is intriguing and is in need of an explanation. In this section, we propose an analytic model in which each emission peak of the I04166 jet corresponds to a pulse of ejection of a spherically expanding wind with an axial density concentration. We show here how the velocity pattern can be naturally explained in this picture. A quantitative description for the slope will be developed, which provides some insight into the model. 
    
\begin{figure}    
\centering
\includegraphics[width=0.47\textwidth]{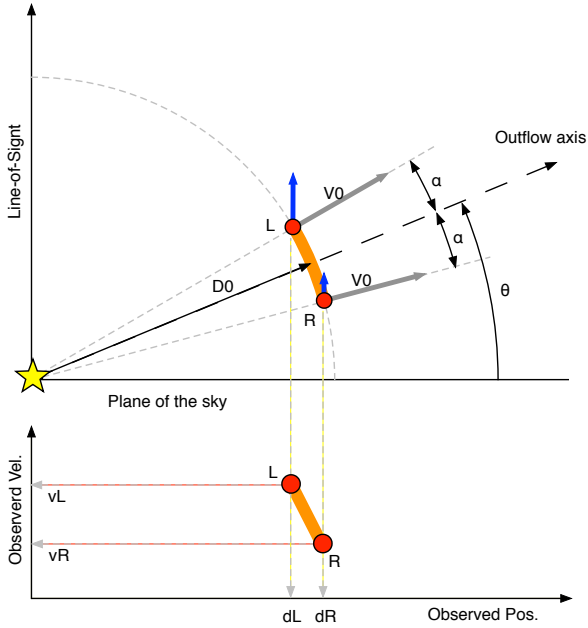}
\caption{An illustration of the axially concentrated spherical wind layer. The upper part illustrates the geometric configuration of the system. The star symbol at the origin labels the position of the driving source, while the thick orange curve represents the ejecta. The gray arrows indicate the intrinsic velocity which is directed only in the radial direction and the blue ones are the line-of-sight component seen by the observer. The lower part of the figure illustrates the corresponding velocity pattern on a PV diagram along the outflow axis. 
}
\label{fig:wind_layer_model}
\end{figure}

    
A cartoon illustration of the model and its corresponding PV diagram is shown in Figure \ref{fig:wind_layer_model}. The star symbol in the upper panel indicates the driving source, and the outflow axis is inclined from the plane of the sky by an angle $\theta$. The ejecta consist of a spherically expanding layer of gas whose velocity is purely in the $r$-direction, pointing away from the origin as indicated by the gray arrows. The mass is only contained within a small opening angle $\alpha$ from the outflow axis, so in three dimensions it will look more like a slightly curved circular plate than a spherical shell. The illustration only shows its cross section in two dimension. In radio observations, an observer (viewing from below) will see a red-shifted lobe in this case.

The lower panel of the same figure illustrates the expected velocity pattern of the model, where a tooth-like pattern is formed. The origin of this quasi-linear velocity gradient can be intuitively understood by considering the gas residing at the left and right edges of the ejecta which we label as $L$ and $R$. In this geometric model, it is apparent that the gas on the left edge will be found at a smaller projected distance from the driving source than the other on the right edge ($d_{L} < d_{R}$). The projected velocity, on the other hand, will be larger for gas at the left edge than that at the right ($v_L > v_R$). This is because the ejecta expand at the designed speed but the flow direction of the left edge is more inline with the line-of-sight direction. Together, this results in an apparent velocity pattern of ``slow head--fast tail'' behavior on the PV diagram. In this picture, the observed velocity gradient is a natural consequence of projection. 
	
For a more quantitative understanding, we characterize a tooth pattern by its slope on the PV diagram by considering the two points $L$ and $R$. For an inclination angle $\theta$ from the plane of the sky, the observable quantities can be written as
\begin{align*}
    	d_L = D_0 \cos(\theta+\alpha),\quad & v_L = V_0 \sin(\theta+\alpha), \\
    	d_R = D_0 \cos(\theta-\alpha),\quad & v_R = V_0 \sin(\theta-\alpha).
\end{align*}
where $D_0$ is the intrinsic distance of the ejecta from the source, $V_0$ is its intrinsic speed, and $\alpha$ is the half opening angle from the axis. The slope $\eta$ of the corresponding velocity pattern is defined as difference in line-of-sight velocity $\Delta v$ over the difference in projected distance $\Delta d$ in the plane of the sky, namely, $\eta\equiv \Delta v/\Delta d $. Considering the two points $L$ and $R$, we have 
\begin{equation*}
    	\eta = \frac{v_L - v_R}{d_L - d_R} = \left(\frac{V_0}{D_0}\right) \frac{\sin(\theta+\alpha) - \sin(\theta-\alpha)}{\cos(\theta+\alpha) -  \cos(\theta-\alpha)}.
\end{equation*}
With some algebra, we arrive at a simple relation: 
\begin{equation} \label{eqn:1}
	    \eta = \left(\frac{V_0}{D_0}\right) \frac{-1}{\tan\theta} .
\end{equation}
Note in the current formulation, a positive $\theta$ value describes a red-shifted gas while a negative $\theta$ corresponds to a blue-shifted one. The apparent position offsets $d_L$ and $d_R$, on the other hand, are considered positive for both cases, which is different from the continuous coordinate in Figure \ref{fig:i04166_obs} where one of the two lobes will be negative. Since $D_0$ and $V_0$ are positive, according to Equation \ref{eqn:1} the slope $\eta$ will be negative for a red-shifted gas and positive otherwise. Note the opening angle $\alpha$ happens to cancel out in this approximation.
    
    
Aside from the outflow inclination angle, Equation \ref{eqn:1} clearly shows that the slope of a tooth feature is determined by the intrinsic velocity of the ejecta and  distance from the driving source. This relation is most interesting when applied to a series of knots in the same outflow. In case of a fixed inclination angle and similar intrinsic speed, the formula predicts that $\eta$ should varies inversely with the distance of knot from the origin. This means that the tooth pattern farther away from the center should present a smaller $|\eta|$ and appear shallower or less steep. This agrees with the trend seen in the I04166 jet. 

\begin{figure}    
\centering
\includegraphics[width=0.47\textwidth]{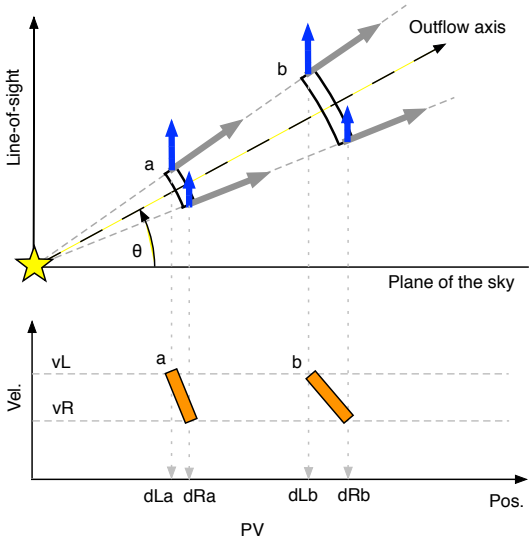}
\caption{Cartoon illustration of geometry of the spherical wind. The upper panel shows the 2D plane containing the outflow axis and the line-of-sight direction. The outflow is inclined from the plane-of the sky by an angle $\theta$. The lower panel shows the corresponding velocity pattern on a PV diagram. Two knots \textit{a} and \textit{b} are plotted, both expanding away from the origin at same constant speed indicated by the gray arrows.  
}
\label{fig:wind_layer_model_two}
\end{figure}

Figure \ref{fig:wind_layer_model_two} intuitively explains why this trend is naturally produced in the current geometric model by comparing two pieces of ejecta $a$ and $b$ at different distances from the origin. For simplicity, we assume that the two ejecta have same intrinsic velocities and sustain same opening angles, which is a good approximation for the I04166 system. In this case, the difference between the line-of-sight velocity components at the left and right edges will be exactly the same for the two ejecta: $|v_L - v_R|$. However, as is shown in the illustration, the outer knot $b$ extends a larger projected distance than the inner one $a$, namely, $|d_{Lb} - d_{Rb}| > |d_{La} - d_{Ra}|$. Hence the velocity pattern of $a$ will turn out to be steeper than that of $b$ because the same velocity difference is distributed across a smaller distance difference in $a$ compared to $b$. Note that the apparent steepness of the pattern depends on the velocity change per unit length ($\Delta v/\Delta d$) and not the actual size of the ejecta. For example, increasing the opening angle of $a$ would result in a larger knot but the pattern on the PV diagram will simply extend and does not become shallower. It would still appear steeper than the outer tooth $b$ regardless of the projected size.  

To more directly describe variation of pattern slope on a PV diagram, we can replace $V_0$ and $D_0$ in Equation \ref{eqn:1} by observable quantities as follows. Let $\bar{v}$ and $\bar{d}$ denote the mean velocity and distance of the two representative points, namely, $\bar{v} \equiv (v_L + v_R)/2$ and $\bar{d} \equiv (d_L + d_R)/2$. We have
\begin{eqnarray*}
    	\bar{v} =& \frac{V_0}{2}(\sin(\theta+\alpha) + \sin(\theta-\alpha)) &= V_0 \sin\theta \cos\alpha\\
    	\bar{d} =& \frac{D_0}{2}(\cos(\theta+\alpha) + \cos(\theta-\alpha)) &= D_0 \cos\theta \cos\alpha . 
\end{eqnarray*}
And therefore,
\begin{equation*}
	\frac{\bar{v}}{\bar{d}} =\left(\frac{V_0}{D_0}\right) \tan\theta.
\end{equation*}
Substituting this into Equation \ref{eqn:1} and eliminate $V_0/D_0$, we derive another expression for the slope:
\begin{equation} \label{eqn:2}
    	\eta = \left(\frac{\bar{v}}{\bar{d}}\right) \frac{-1}{\tan^2\theta}.
\end{equation} 
    
Equation \ref{eqn:2} expresses $\eta$ in terms of the mean projected velocity and mean projected distance of the two points $L$ and $R$. This explicitly shows that the slopes of the ejecta should vary inversely with their position offset from the central source for a same inclination angle and similar mean observed velocity $\bar{v}$.  
    
\begin{figure*} 
\includegraphics[width=\textwidth]{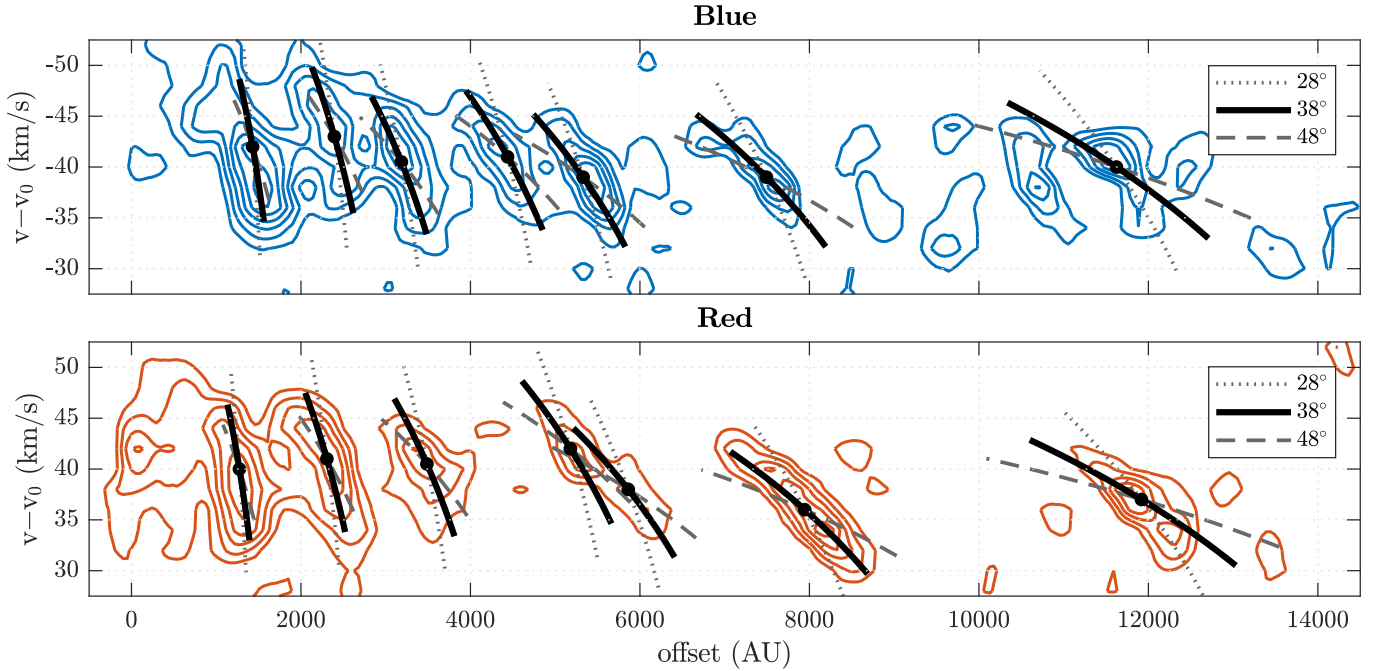}
\caption{\co{2}{1} PV diagram in the EHV regime shows the sawtooth-like pattern of the I04166 jet overlaid by model predictions. The contours start from $0.4$ K and the step is $0.25$ K. Patterns predicted by the concentrated spherical wind model at three assumed angles from the plane of the sky are shown. The solid lines are results assuming that the outflow axis is $38$\arcdegsp inclined from the plane of the sky (an inclination angle of $52$\arcdeg), which fits the overall pattern reasonably. The dotted and dashed lines are predictions assuming $28$\arcdegsp and $48$\arcdegsp from the plane of the sky and they deviate more from the observation. Each line segment represents an ejected knot of $7.5$\arcdegsp half opening angle (see Figure \ref{fig:wind_layer_model}), and the black dot in the middle mark the center of the knot. 
}
\label{fig:i04166_co21_pvmodel}
\end{figure*}

  	
This prediction can be tested with the I04166 jet. Figure \ref{fig:i04166_co21_pvmodel} shows a comparison between model and the observational data. At each knot position, we overlay three sets of velocity pattern computed with different inclination angles on top of the observed \co{2}{1} data. Given an inclination angle, we can compute the intrinsic distance $D_0$ and the intrinsic speed $V_0$ of each ejected blob from its observed position and velocity. And with the values of $D_0, V_0$ and $\theta$ the velocity pattern is decided. In this demonstration we display the model pattern up to a half-opening angle of $\alpha=7.5$\arcdegsp from the rough center of each tooth.

The thick dark lines show the resulting pattern of the model that assumes an inclination angle of $38$\arcdegsp from the plane of the sky ($52$\arcdegsp from the line-of-sight). This is the average angle estimated by \citet{tafalla2017} based on the elliptical shapes of the knots B6 and R6 in their ALMA observations. Figure \ref{fig:i04166_co21_pvmodel} shows that such angle is not only suitable for predicting the pattern of these two particular knots, but also results in good fits for most others. On the other hand, the patterns computed with $\pm10$\arcdegsp from this value are less compatible with the observation. The predicted slope is too steep in the case of $28$\arcdegsp and too shallow in the case of $48$\arcdegsp. The overall behavior of the whole sawtooth velocity pattern can therefore be generally explained with one same inclination angle of $\simeq38$\arcdegsp from the plane of the sky. Since the model dictates that the pattern slope should vary inversely with knot distance from the center (Equation \ref{eqn:2}), the simultaneous agreement across the knots is actually not trivial. The overall agreement implies that the proposed geometric model does capture the feature of the I04166 jet kinematics. 
	
Figure \ref{fig:i04166_co21_pvmodel} also shows that the model velocity pattern are not exactly linear. They are slightly curved and this can be understood with Equation \ref{eqn:1}. Consider a single dark line segment computed for an ejected blob of fixed $D_0$ and $V_0$ with an inclination angle of $\theta=38$\arcdegsp from the plane of the sky. Although $D_0$ and $V_0$ are the same for different points along this stripe, the angle of $38$\arcdegsp does not apply to the entire piece. As we have assumed a half opening angle of $7.5$\arcdegsp, the upper and lower ends of the model tooth pattern on the PV diagrams are actually inclined by $45.5$\arcdegsp and $30.5$\arcdegsp, respectively. According to Equation \ref{eqn:1}, $|\eta|$ should vary inversely with $\tan\theta$ for a fixed $V_0/D_0$. Therefore at the lower end of the velocity pattern, $|\eta|$ is always $\sim1.73$ times larger than that at the upper end in this example. Because the precise inclination angles of different parts of an ejected blob are different, the pattern appears curved. 
    
Finally, we go one step further and absorb the term $\bar{v}/\bar{d}$ into $\eta$. A dimensionless slope $\hat{\eta}$ can then be defined as
\begin{equation*} 
    	\hat{\eta}\equiv \frac{\eta}{\bar{v} / \bar{d}} 
        =\frac{\Delta v / \bar{v}}{\Delta d / \bar{d}} 
    	=\frac{(v_L - v_R)/(v_L + v_R)}{(d_L - d_R)/(d_L + d_R)}. 
\end{equation*}
From Equation \ref{eqn:2}, we have
\begin{equation} \label{eqn:3}
	    \hat{\eta} = \frac{-1}{\tan^2\theta} .
\end{equation}
The dimensionless slope is simply a function of the outflow inclination angle $\theta$ from the plane of the sky. Its value is negative for both the red and blue-shifted gas.
	
Since $\hat{\eta}$ is an observable quantity that can be calculated from the observed tooth-like velocity pattern, Equation \ref{eqn:3} actually provides a way to estimate the outflow inclination angle. For any two chosen points on a tooth pattern, it gives you the average inclination angle of the two specific points (see Figure \ref{fig:wind_layer_model}). To estimate the inclination angle of the outflow axis, we therefore need to have $L$ and $R$ symmetrically displaced from the outflow axis by a same half-opening angle. Another caveat is that this method only applies when the pattern forms by a spherical wind-like velocity field as assumed. It will not just apply to any linear velocity pattern on a PV diagram.
	
\begin{figure}    
\includegraphics[width=0.47\textwidth]{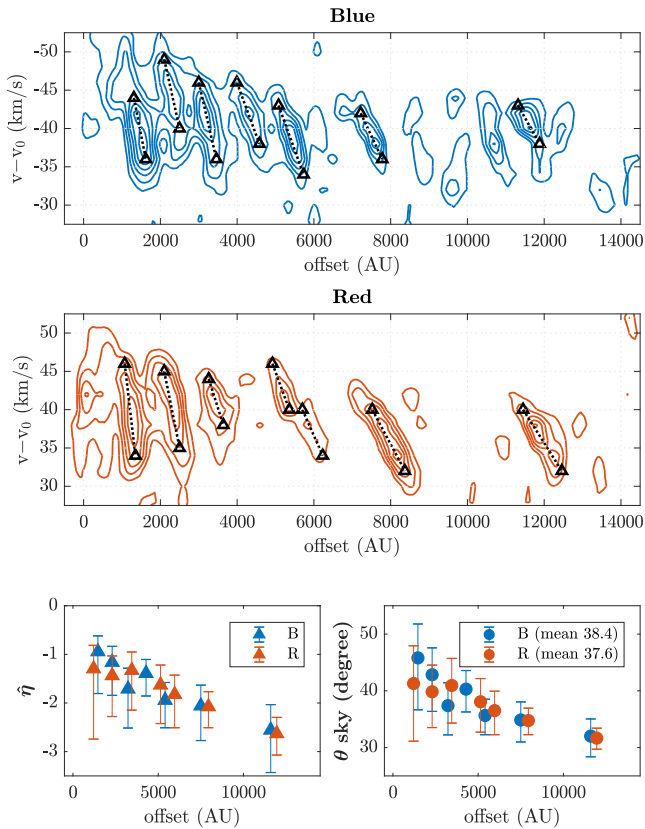}
\caption{Estimating inclination from the observed dimensionless slope $\hat{\eta}$ using Equation \ref{eqn:3}. The triangle symbols in the top two panel label the two points used to estimate $\hat{\eta}$. In the bottom left and right panels plotted are the obtained $\hat{\eta}$ values and the inferred inclination angle from the plane of the sky $\theta$ as a function of mean projected distance of each tooth pattern. The error bars are estimated by displacing the upper triangle of each pattern by $\pm140$ AU to display the uncertainties.
}
\label{fig:estimate_inclination}
\end{figure}

To see how this method works in practice, we apply it to the pattern of I04166. We manually pick out two points for each tooth feature as shown in the upper two panels of Figure \ref{fig:estimate_inclination} where the selected points are labeled with open triangles. The corresponding dimensionless slopes are computed for each tooth and plotted on the bottom left panel. The value of $\hat{\eta}$ falls in the range of $\sim-3$ to $-1$. The bottom right panel plots the corresponding inclination angles computed using Equation \ref{eqn:3}. The mean inclination angle values are $38.4$\arcdegsp and $37.6$\arcdegsp for the blue and red shifted lobes, respectively, which seems to be a reasonable estimate. Note that there is a trend where $\hat{\eta}$ (and $\theta$) appear to decrease with distance from the source which is not really expected.

Given the finite resolution of observations, we might not always be able to choose perfect representative points for a tooth feature that best reveal its inclination. To assess the uncertainty associated with this potential error, we displace the upper selected point in each pattern to the left and right by 140 AU ($\sim$1\arcsec\, at 140 pc distance) and compute the associated $\hat{\eta}$ and $\theta$ again. The varied values are then presented as the upper and lower bounds of the error bars in the panels of $\hat{\eta}$ and $\theta$ in Figure \ref{fig:estimate_inclination} to illustrate the associated error range. The larger error bars for the inner data points indicate that they are more sensitive to the position measurement. This is because the inner knots generally present a steeper velocity pattern and a smaller $\Delta d$. Therefore a same positional error would cause a more significant change in the estimated $\theta$. Considering the size of the error bar, the inferred $\theta$ from most of the knots appear to be consistent with the mean value of $\sim38$\arcdegsp. The decreasing inferred $\theta$ from inner to outer knots, however, is puzzling as it appears too systematic to be random measurement errors. Since it is unlikely that the inclination angle of the I04166 outflow has really changed by $15$\arcdegsp over the years, this probably suggests that there are some missing elements in the current simple model. One potentially possible cause of this behavior is mentioned in the next section.

\section{Numerical Simulations of Variable Velocity Wind}
		\label{sec:numerical_simulation}
	
The kinematics of the I04166 molecular jet is compatible with a spherical wind of axial density concentration. We will further demonstrate the scenario by reproducing such velocity patterns using numerical simulations. Our simulations build upon the unified wind model framework outlined in \citet{shang2006} with detailed simulation setups following the two-temperature generalization described in \citet{wang2015}.

\subsection{The Unified Wind Model for Young Protostellar Outflows}
	
In the framework of the unified wind model, a toroidally magnetized wind is launched into an ambient envelop of toroid. The primary wind is constructed based on the asymptotic behavior of the $X$-wind \citep{shu2000,shu1995}, which is characterized by a radially directed velocity field and cylindrically stratified density profile that is highly peaked toward the axis ($\rho\propto\varpi^{-2}$, where $\varpi$ is the cylindrical distance from the outflow axis). Meanwhile, the ambient mass distribution is described by the singular isothermal toroid solutions of \citet{li1996}. Its density is the highest near the equator and decreases to nearly zero in the polar regions. Combining the wind and toroids, the unified wind model can naturally explain the dual high-velocity jet and low-velocity shell components commonly observed in the youngest group of molecular outflows associated with the Class 0 protostars. 
	
Using 2D axisymmetric simulations, \citet{shang2006} shows that the dense axial region of the primary wind can form a collimated jet-like structure along the axis, and the diverging part of the wide-angle wind will gradually sweep up the ambient mass to form a cavity-like structure near systemic velocity. The follow-up work of \citep{wang2015} decouples the wind temperature from the ambient by using a wind tracer field. This two-temperature scheme allows the study of interplay between wind magnetization and thermal pressure. 
	
To more realistically compare the simulation and observations, we construct synthetic maps and PV diagrams for the simulations. In estimating the \co{2}{1} emission lines, we calculate the level population of different rotational $J$ states assuming statistical equilibrium between processes of spontaneous emission and collision (de-)excitation with molecular hydrogen. For simplicity, effects of radiative excitation and de-excitation on the level populations were ignored so the line emissions can be calculated locally. Such effects may be more important for line ratio studies, but should have little impact on the gas kinematics we are interested in here. The main difference between a statistical equilibrium result and a thermal population is found at lower density regimes. When collisional processes are not frequent enough to thermalize the population, the lower rotational states will be relatively more occupied. To obtain image cubes, we perform ray tracing along line-of-sight direction and assume a line width of 1 km\,s$^{-1}$ when accounting for the gas velocity.

The primary wind in the unified wind model actually bears similarities with the picture illustrated in Figure \ref{fig:i04166_co21_pvmodel}. Its wind velocity field is also directed along spherical-$r$ direction and has an angle-independent magnitude. Its density profile is cylindrically stratified and is highly concentrated toward the axis, which effectively makes the wind only visible close to the axis, resembling the finite size (opening angle) of the ejecta. Because of the similar setup, it is quite straightforward to apply the simulations to our current problem. Our goal is to reproduce the observed kinematic feature of the I04166 jet, so we will only focus on the high-velocity feature from the axially concentrated part of the primary wind. A few modifications are adopted to better fit the observations of I04166.
	
The observed knot width of the I04166 jet is better reproduced when we include an inner cone for the wind. Within $3$\arcdegsp half-opening angle, the rising density and toroidal magnetic field strength is capped at constant values. The gas behaves and interacts differently according to the toroidal magnetic field strength.  The tooth-like velocity pattern is better maintained for weaker magnetization. We adopt a large Alfv\'en Mach number of $M_A=150$ for purposes of better modeling. The gas kinematic temperature is assumed to be $100$\,K.

\subsection{Modeling the Ejection History}
	
An ejection history needs to be designed to reproduce the observed velocity pattern. In the model framework, this corresponds to specifying a velocity and density history for the wind. We adopt an inclination angle of $52$\arcdegsp (or $38$\arcdegsp from the plane of the sky), and make a few assumptions to simplify this task. The first is that we fix the wind density throughout a simulation and only vary the wind velocity. And the second is that we will try to follow a periodicity of $\sim93$ year, which is estimated from the roughly equal spacing of the first few knots of I04166 jet. The dynamical age of the fitted regions (up to knots B7 and R7) is about 1200 years.


In the simplest of possible scenarios, we can model each tooth pattern using a single burst. This also most closely resembles the picture discussed in the previous section. Periodic bursts in protostars may occur through episodic accretion, which could arise, for example, in a gravitationally unstable circumstellar disk disturbed by binary companions. Evidence of submillimeter variability has been reported for the Class I protostar EC 53 in Serpens Main \citep{hodapp2012, yoo2017}. Theoretically, material can also build up in a disk before it rapidly accretes and results in a burst. Details of the processes are less well understood especially in the protostellar phase \citep{hartmann2016}. Regardless of the origins, as the ejections come from physically small regions close to the star, the instantaneous perturbation to the velocity can be considered constant from the view of modeling purposes.

\begin{figure}    
		\centering
		\includegraphics[width=0.47\textwidth]{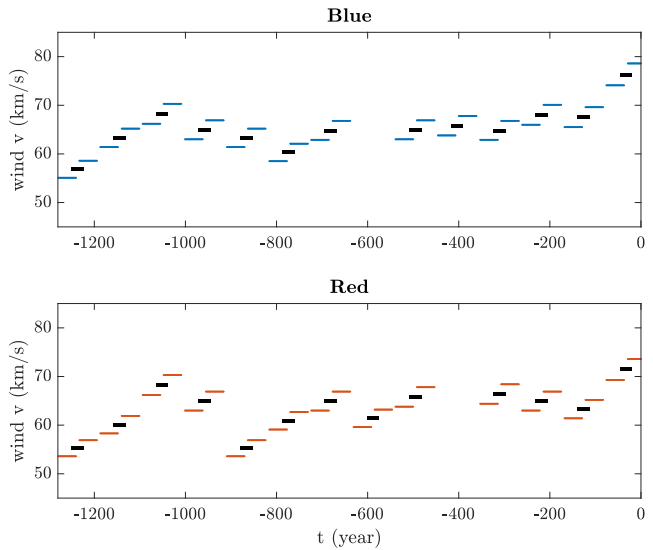}
		\caption{Velocity history model behind simulations that generate the PV diagrams in Figure \ref{fig:sim_pv_single_burst} and \ref{fig:sim_pv_cycle}. The horizontal axis shows the simulation time where $t=0$ represents the time when outflow is observed. The shorter dark segments are the single burst model while the longer blue and red lines are the low--high cycle model.   
		} 
		\label{fig:velocity_history}
	\end{figure}

Figure \ref{fig:velocity_history} shows the model velocity history in which $t=0$ represents the time when the outflow is observed. The thicker dark line segments mark the time and speed of the bursts. They are separated by $\sim93$ years in time, each having a constant speed and lasts for $20$ years long. The velocities of the ejecta are designed so as to let them travel ballistically to the observed location by the time of $t=0$. Between the bursts, the velocity is zero so no material is ejected during the break. Note that the simulations are intended to explore the feasibility of the concentrated spherical wind geometry, so the velocity histories shown here fits to the history of I04166 outflow. For example, we have actually skipped one cycle in both the blue and red-shifted lobes to better match the observed pattern. 
	
\begin{figure}    
\includegraphics[width=0.48\textwidth]{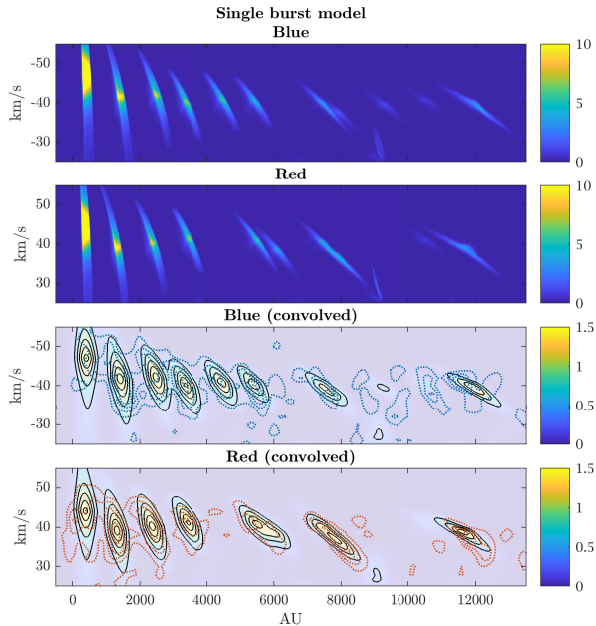}
\caption{Numerical simulations reproducing the observed velocity pattern of I04166 assuming multiple $20$ year long bursts occurring on a $\sim93$ year period. Each observed tooth pattern either results from a single ejecta or multiple collided ones. A gas kinetic temperature of $100$\,K is assumed while computing the synthetic line emissions. The upper two panels show the PV diagrams cut along the outflow axis of the blue and red-shifted lobes. The lower two panels are the same results after first convolving the datacube with a Gaussian beam of 420\,AU full width at half maximum (3\arcsec\, at 140 pc). The first contour is $0.4$ K and the step is $0.35$ K.  
}
\label{fig:sim_pv_single_burst}
\end{figure}

Figure \ref{fig:sim_pv_single_burst} shows the resulting synthetic PV diagrams along the outflow axis. They clearly present a series of linear features similar to the analytic results in Figure \ref{fig:i04166_co21_pvmodel}. The slope of the pattern varies with offset of distance, where the inner ones are apparently steeper than the outer ones. The simulation therefore show that the velocity pattern of spherical wind-like ejecta is indeed like a sawtooth. 
For a realistic comparison with observation, we also convolve the synthetic image cube with a Gaussian beam of $3$\arcsec\, full width at half maximum to mimic the resolution of the \co{2}{1} data \citep{santiago-garcia2009}. The lower two panels compares the convolved data (dark contours) with observations (dotted color contours). Although not perfectly, our numerical simulations successfully reproduce the observed velocity pattern.

While things seem to work out of the box, some tweaks on the velocity history are actually needed to achieve this desired outcome. The main problem is that the emission of an ejecta will gradually become weaker as it travels away from the origin. This is because its structure is becoming more diffused due to both the expansion and the effect of gas pressure. To achieve a comparable emission to the observations, we have put multiple ejecta together to form the knots beyond $\sim7000$ AU. We do this by arranging a $\sim3-5$ km\,s$^{-1}$ difference in intrinsic velocity between successive periods so that they end up running into each others to form a more massive feature. This is, of course, not the only way of achieving this. Directly raising the wind density for the particular knots may also work. It is probably not easy to tell which approach is closer to reality with the current data. 
	
	
While the single burst design can successfully produce the main velocity structure, the generated tooth patterns (Figure \ref{fig:sim_pv_single_burst}) appear to be too isolated from each other compared to the observations. This may suggest that the true velocity history is more continuous than just a series of distinct bursts. As knots are commonly formed through the catching up process in a variable velocity flow, we now try to model the same pattern with a more continuous velocity history that consists of many low--high velocity cycles similar to a square wave. The revised velocity history is shown with blue and red line segments in Figure \ref{fig:velocity_history}. 
A $40$ year higher velocity episode now follows a $40$ year lower velocity episode preceding it within each $\sim93$ year period. Nothing is launched during the remaining gaps. The mean velocity of each cycle is similar to the velocity in the single burst case, and the difference between high and low velocities corresponds to a wave height of $0.03$. In this design, the higher velocity material is expected to run into previous lower velocity material, generating a denser layer in between. 
	
\begin{figure}    
\includegraphics[width=0.47\textwidth]{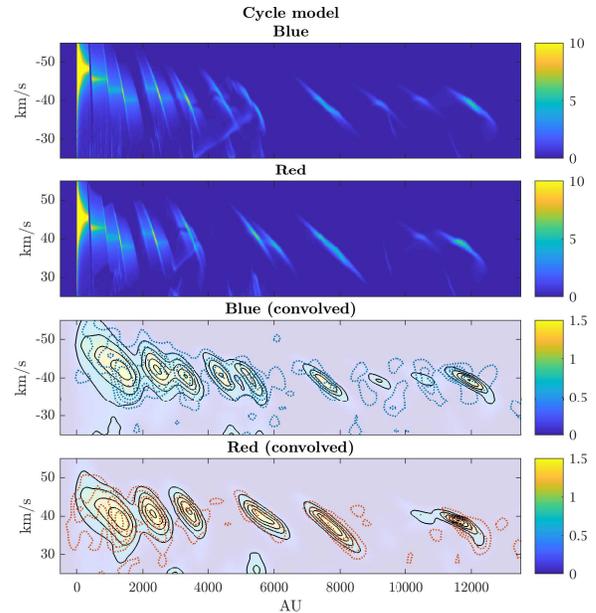}
\caption{Numerical simulations reproducing the observed velocity pattern of I04166 by assuming a more continuous ejection history of low--high velocity cycle. 
Within each $\sim93$ year period, $40$ years of higher velocity ejection follows the lower velocity ejection.  Plot details are the same as Figure \ref{fig:sim_pv_single_burst}.
	}
	\label{fig:sim_pv_cycle}
\end{figure}
	
	The PV diagrams from the low-high cycle approach is presented in Figure \ref{fig:sim_pv_cycle}. The new velocity history produces sawtooth pattern at large distance just like the single burst model, but the pattern is more complex in the inner region. Although the materials ejected during the low and high velocity episodes of a cycle are initially distinct, they gradually merge together and move ballistically as one single packet. In the current case, they appear as a single feature after a few thousand AU projected distance from the origin.
	
	After convolution, the detailed velocity history are no longer discernible in the PV diagrams, but the differences in velocity history do affect the appearance of the tooth pattern especially for the inner ones (the lower two panels of Figure \ref{fig:sim_pv_cycle}). The first difference is that the tooth now appear slightly wider. This is because the outflow mass is now more continuously distributed along the outflow axis instead of being only confined in discrete ejecta right from the beginning. The second difference is that the apparent slope of the tooth pattern is modified. This is because we are now not just observing the denser layer of mass. Emissions from the leading low-velocity gas and the trailing high-velocity gas also contribute to the velocity pattern, and they modify the tooth in a way that makes it look less steep. 
	
	Comparing the results in Figure \ref{fig:sim_pv_single_burst} and \ref{fig:sim_pv_cycle}, we see that the more evolved outer knots are indeed well described by the model of thin spherical shell. But for the younger knots closer to the driving source, one needs to worry about the continuous velocity variation as the gas has not yet evolved or collided into a thin layer. Interestingly, this effect may potentially explain the systematically smaller dimensionless slope magnitude $|\hat{\eta}|$ of the inner knots compared to the more evolved outer ones (Figure \ref{fig:estimate_inclination}).

\begin{figure}    
		\includegraphics[width=0.47\textwidth]{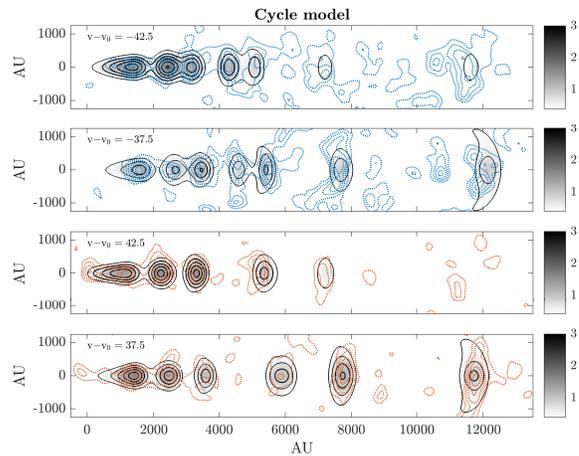}
		\caption{Synthetic \co{2}{1} channel maps of the cycle history model simulations that produces the PV diagram in Figure \ref{fig:sim_pv_cycle}. Emission is averaged over a 5 km\,s$^{-1}$ bin in each channel. The gray scale and the dark contours show the model emission. The red and blue contours show the I04166 observational data.	The simulation result have been convolved with a Gaussian beam of 420\, AU FWHM (3\arcsec\, at 140 pc) before making the PV diagram to mimic the resolution of the observational data. The first contour is $0.4$ K and the contour step is $0.25$ K.
		} 
		\label{fig:simulation_channel_map}
	\end{figure}

	
	Finally, we check how the synthetic channel maps compare to the observations for completeness (Figure \ref{fig:simulation_channel_map}). Four channels centering at $\pm37.5$ and $\pm42.5$ km\,s$^{-1}$ relative to the system velocity are presented to compare the knot morphology of the EHV gas. 
	The simulated knot positions generally agree with the observation and the strength are also similar. The most apparent differences are found in the size of the knot especially in the blue shifted lobe. For the outermost blue knot (B7) in the top panel and the second outermost blue knot (B6) around $\sim7900$ AU in the second panel, the observational data are clearly more extended than the simulation result. For the knot B7, we did not really attempt to reproduce its emission because its velocity pattern along the axis is quite fragmented in the first place. It is therefore not surprising that the synthetic channel maps do not match well there. For the knot B6, it turns out that its relatively short tooth feature on the PV diagram we have successfully reproduced is misleading about its lateral size. The channel map shows that it is actually quite extended in the lateral direction. The asymmetric knot is not well represented by the current axisymmetric model.

	Numerical simulations in framework of the unified wind model can successfully reproduce the sawtooth-like velocity pattern observed in the I04166 molecular jet. This supports the proposed spherical wind geometry interpretation of the pattern.

\section{Discussion}

	\label{sec:discussion}
	
\subsection{Wind or Jet}
	
	Historically, models for outflows fall into two main paradigms: of wide angle wind and collimated jet. The former can explain the wide opening of outflow cavities while the latter fits the behavior of more collimated systems. In Class 0 outflows, both jet and shell components are simultaneous present and the unified wind model provides a more natural explanation. In this picture, the outflow is a spherically directed wide angle wind and the central jet-like feature arise from the high density concentration near the axis. The I04166 molecular outflow has been found to present low and high-velocity emissions that form spatially distinct components. The much wider opening of the low-velocity shell suggests it is not driven by the collimated central jet component, and probably represents swept-up material from some underlying wide angle wind that is not directly visible \citep{santiago-garcia2009}. The analysis of this work shows that the ordered sawtooth velocity pattern seen in the jet component can actually represent a spherical wind-like velocity field, which further reveal the wind nature of the source. The observations of the I04166 outflow is therefore in support of the picture outlined in the unified wind model of \citet{shang2006}, in which a magnetocetrifugal wind is launched into an ambient toroid.
	
	
	While a sawtooth velocity pattern is well explained by this wind picture, an authentic jet paradigm can actually also generating a similar pattern. The key to have lateral velocity component from a parallel jet is to have post-shock gas ejected from an internal working surface arising from a variable velocity flow. \citet{santiago-garcia2009} has pointed out that the numerical simulations of pulsed jets in \citet{stone1993b} can result in a sawtooth-like velocity features similar to that observed in the I04166 jet. The kinematics of the two extended knots R6 and B6 (near $\sim7900$ AU offsets from the source) mapped by ALMA in are also explained by the geometric model proposed in \citet{tafalla2017}. The authors found that, after subtracting out the mean motion of the moving knots, the residual velocity field can be well described by a slightly curved expanding disk with an expansion speed increasing linearly from $0$ km\,s$^{-1}$ at the center to $13$ km\,s$^{-1}$ at the edge, and such expansion velocity may be potentially attributed to side-way ejection from an internal working surface.

\begin{figure}    
	    \centering
		\includegraphics[width=0.48\textwidth,trim={0 1cm 0 1cm},clip]{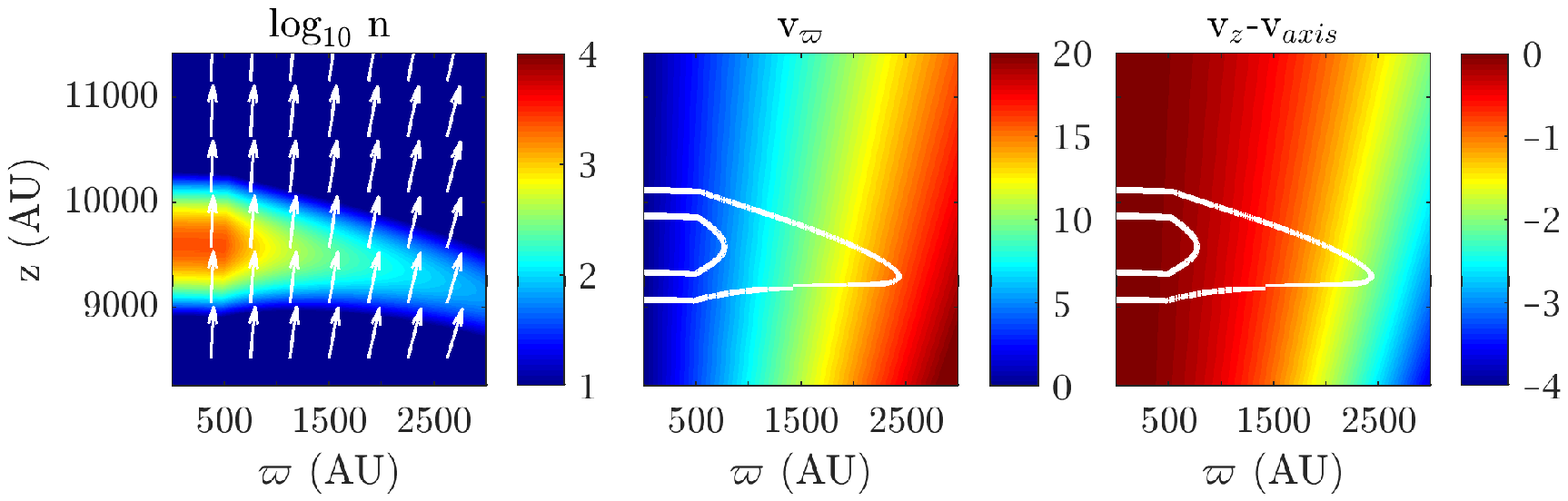}
		\includegraphics[width=0.48\textwidth,trim={0 1cm 0 1cm},clip]{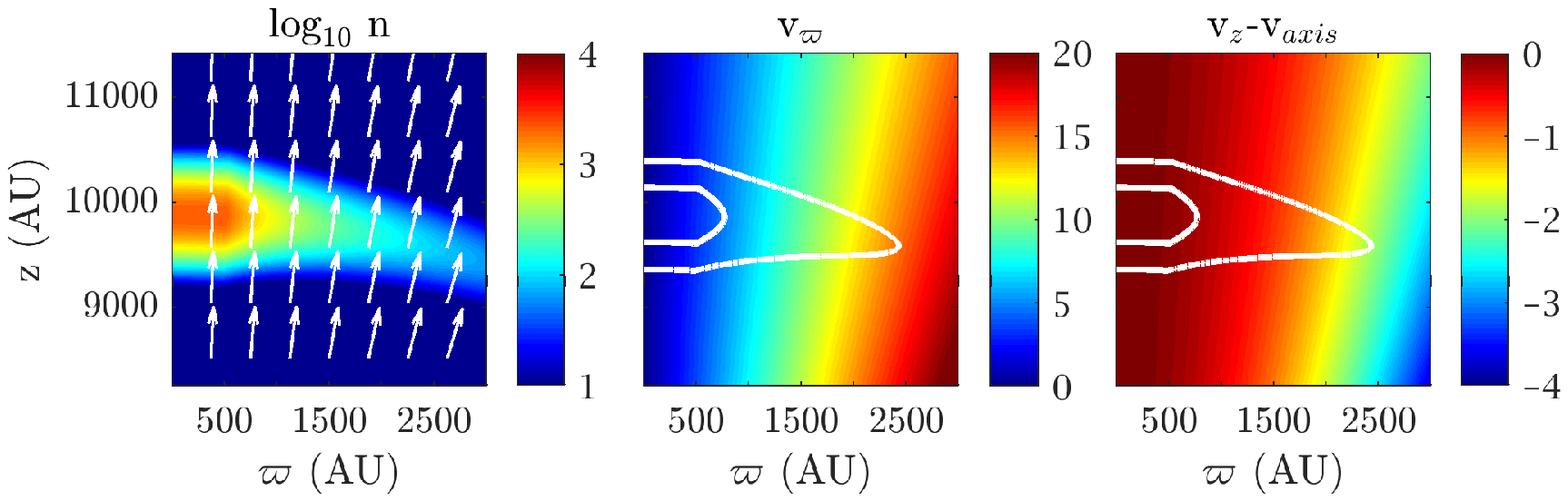}		
		\caption{Meridianal density and velocity structure of a axisymmetric model for knots B6 (\textit{upper}) and R6 (\textit{lower}). The first column shows the number density on a logarithmic scale, and the white arrows indicate the velocity direction. The second column shows the velocity component perpendicular to the outflow axis ($v_\varpi$). The third shows the velocity component parallel to the outflow axis after subtracting the gas velocity on the axis ($v_z - v_{axis}$). 
		The white lines are isodensity contours at $10^2$ and $10^3$ cm$^{-3}$ to guide the eyes.
		} 
		\label{fig:model_knot6_2d}
	\end{figure}

\begin{figure*}    
	    \centering
		\includegraphics[width=0.95\textwidth,trim={0 1cm 0 1cm},clip]{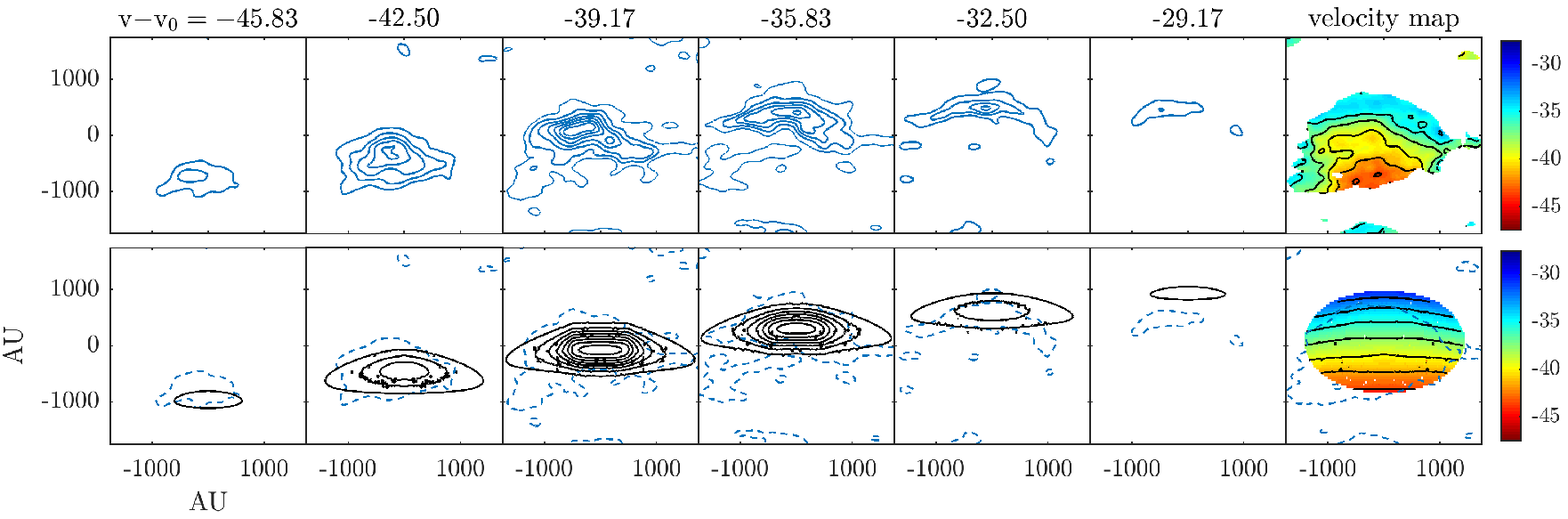}
		\includegraphics[width=0.95\textwidth,trim={0 1cm 0 1cm},clip]{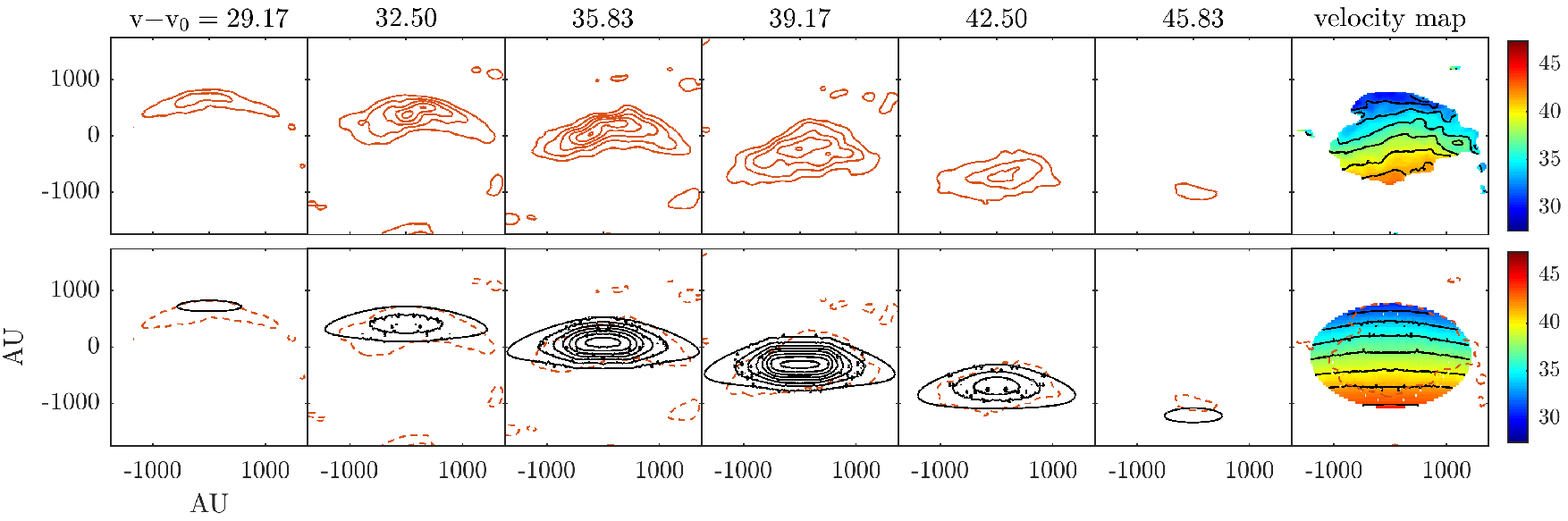}
		\caption{Comparison between model and ALMA \co{2}{1} observations of knots B6 (\textit{upper}) and R6 (\textit{lower}). The observational data are from \citet{tafalla2017}. For each lobe we show observational data in the upper row and the model result in the lower. The last column is the velocity map (only regions with integrated emission above a threshold are shown), while the rest are channel maps. The emission contours starts from $0.15$ K and the step is $0.25$ K. The step of the velocity contour is 2 km\,s$^{-1}$. 
		The first emission contour of the observational data are also plotted in the model panels with dashed lines for a better comparison.
		} 
		\label{fig:model_knot6_chmap_m1}
	\end{figure*}

	It turns out that the kinematics of the model proposed in this work can appear quite similar to the model of \citet{tafalla2017}. This becomes apparent if we also switch our reference frame to one where the tip of the blob is at rest. Consider an arbitrary part of the ejectum that is at an angular offset of $\alpha$ from the outflow axis. Its distance from the center along the surface is proportional to $\alpha$, and its perpendicular velocity component is proportional to $\sin\alpha$. For a small angle, $\sin\alpha$ is approximately equal to $\alpha$, which means that the perpendicular (or ``side-way'') velocity component will increase linearly with distance from the axis, similar to the picture of an expanding disk discussed in \citet{tafalla2017}. To explicitly show that the high resolution ALMA data of knots B6 and R6 can be explained by both the expansion disk scenario and our spherical wind scenario, we construct a simple 2D axisymmetric model for knot B6 and R6 with the meridianal velocity distributions specified by a spherical wind. The model density and velocity distributions are shown in Figure \ref{fig:model_knot6_2d}, and the corresponding synthetic images are presented in Figure \ref{fig:model_knot6_chmap_m1}.
	
    The upper and lower rows of Figure \ref{fig:model_knot6_2d} are models for knots B6 and R6, respectively. We use exactly the same velocity field and they differ only in the knot position from the origin. We mimic the setup of the numerical simulations presented in Section \ref{sec:numerical_simulation} so the wind is purely in the spherical-$r$ direction and of a constant magnitude. The density profile drops off with distance from the axis as $\rho\propto\varpi^{-2}$, but is constant inside an inner cone of $3$\arcdegsp half opening angle. To model the density distribution of a thin mass layer, a Gaussian profile of FWHM $490$ AU ($3.5$\arcsec) centering at a fixed distance from the driving source at the origin is applied along spherical-$r$ direction. 
    The magnitude of the perpendicular velocity component is determined by the distance of the ejecta from the driving source (and the adopted inclination angle). Its magnitude increases roughly linearly from 0 to $\sim15$ km\,s$^{-1}$ from the axis to the outer edge of the white contour as shown in the second columns of the figure. 

    Figure \ref{fig:model_knot6_chmap_m1} compares the synthesized model channel maps and velocity maps for the knot R6 and B6 to the observations. The knot positions within each channel and the characteristic velocity gradient along the axis are well reproduced. The main difference is that the observed emissions appear to be more fragmented than the model, especially in the blue lobe. The observations also appear to be relatively more curved than the model emission, which may suggest that the ejecta are slightly more curved than those from spherical shell.
    
	While both scenarios can similarly explain a tooth-like velocity pattern (because of their approximately equivalent apparent kinematics), the spherical wind model is more constraining especially when applied to a series of knots. It predicts that the dimensionless slope $\hat{\eta}$ of all tooth features should all be the same and only depend on the outflow inclination angle (Equation \ref{eqn:3}). As illustrated in Figure \ref{fig:i04166_co21_pvmodel}, when the projected velocities and offsets are both fixed in a given observation, the outflow inclination angle is the only free parameter that can be varied to fit the sawtooth velocity pattern. The reasonable agreement across the entire sawtooth pattern is therefore a compelling sign of an underlying wind-like geometry. On the other hand, in the paradigm of parallel jets with internal working surfaces, the slope of the tooth feature should depend critically on the strength of the side-way ejection velocity (besides the jet inclination angle) because this is the cause of the apparent velocity gradient on the PV diagram. As the shock conditions may simply differ for different knots, there is no guarantee that the series of teeth pattern should behave coherently, and the observed systematic variation of slope with distance is not automatically explained in this paradigm. 
	In other words, the origin of lateral velocity components in a spherical wind scenario is purely geometrical, so it is not possible to vary the predicted slope of some of the teeth while keeping others unchanged. The systematic and coherent behavior of the teeth slopes in the I04166 jet is therefore a diagnostic of a spherical wind-like velocity field, although this does not rule out the possibility of a jet shock. We note that a thermal sound speed of $\sim 15$ km\,s$^{-1}$ corresponds to a thermal temperature of $>60000$ K. A jet-shock scenario would therefore predict very high temperature at least for the freshly-shocked inner knots before heavy cooling can take place.

\subsection{Implications on Velocity History}
	
	
	A correct geometric picture is crucial for inferring the intrinsic outflow velocities from the observed line-of-sight components, which is fundamental for the study of outflow velocity history. For example, observations of the I04166 jet reveal an apparent velocity variation on order of $\sim5$--$10$ km\,s$^{-1}$ within each emission peak. Without any prior knowledge, one might simply de-project the velocity using a fixed inclination angle and conclude that the flow velocity varies periodically with a wave height of $\sim8$--$16$ km\,s$^{-1}$. The catch of this seemingly natural conclusion is that it implicitly assumes a geometric picture of unidirectional flow with varying speed along the axis, which is probably not true. The modeling of \citet{tafalla2017} showed that a tooth pattern likely results from the lateral velocity of the flow, and in this work we show that the entire pattern can be consistently explained by a spherical wind like geometry. This leads to a very different conclusion where the observed  $\sim5$--$10$ km\,s$^{-1}$ apparent velocity variation in each peak may have little to do with outflow velocity variation. As demonstrated in the simulation of the single-burst velocity history model (Figure \ref{fig:sim_pv_single_burst}), ejecta of constant speeds can reproduce the observed feature well.

    A ``slow head--fast tail'' velocity pattern on a PV diagram could in principal form in three ways. The first is a diverging velocity field like the spherical wind discussed in this work. The second is the side-way ejected gas from internal working surfaces of shocks. The third is an intrinsic velocity variation along the flow direction, which is also the true velocity history of general interests. In the particular case of I04166 outflow, the series of clearly detected tooth-like pattern have allowed us to reveal its wind nature. In a more general case without multiple clearly-detected teeth spanning a range of distance from the origin, however, discriminating between the potential origins is difficult. The three mechanisms are also not mutually exclusive, and could be all at work in a same source. A long as the flow velocity field is diverging in nature, there will be a corresponding dispersion in the observed line-of-sight velocity component due to projection. As long as there is compressed, high density region, lateral velocity could be generated from the pressure gradient force like in the case of an internal working surface, and the real question is whether it is significant or negligible. And as long as the outflow launching velocity is time dependent, there will be velocity gradient along the flow direction, whether it is an unidirectional jet or a diverging wind. Disentangling contributions from these different mechanisms is therefore required for an accurate understanding.

\begin{figure}    
		\centering
		\includegraphics[width=0.47\textwidth]{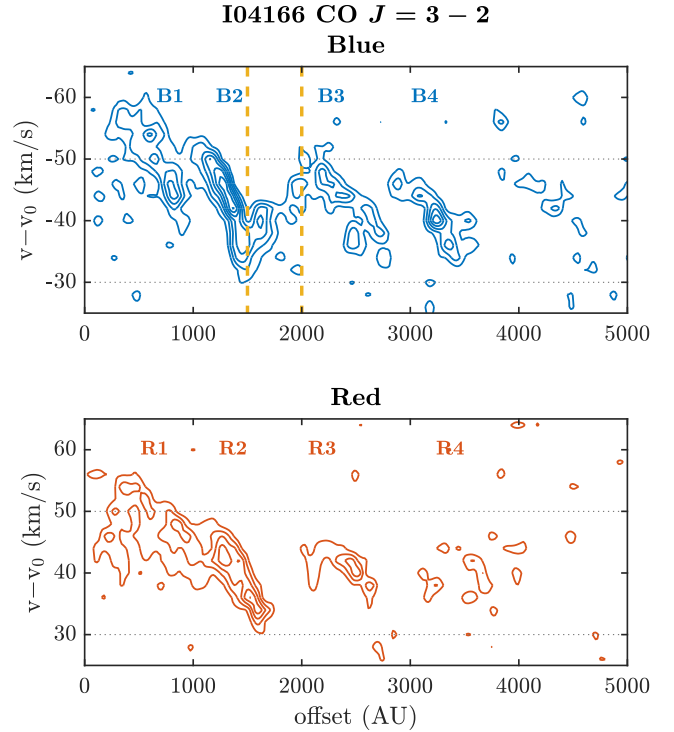}
		\caption{\co{3}{2} PV diagrams of the I04166 jet in the EHV regime replotted from \citet{wang2014}. The velocity pattern is more complicated than predicted by the simple spherical wind model. The pattern between two vertical dashed lines, for example, are not explained by the current model. These youngest gas at regions closest to the central source may reveal clues of the detailed outflow ejection history.
		} 
		\label{fig:i04166_obs_co32}
	\end{figure}

    The velocity record is better preserved in regions closer to the driving source before gas traveling at different speeds collides with each other and merges. For example, the resulting synthetic PVs of the model of single burst history in Figure \ref{fig:sim_pv_single_burst} and the model of more continuous cycle history in Figure \ref{fig:sim_pv_cycle} are notably different within a few thousand AU but appear almost identical beyond that. High angular resolution observations toward the inner region of the outflow is needed to assess the velocity history. For example, \citet{wang2014} mapped the inner few thousand AU of the I04166 jet in \co{3}{2} using the Submillimeter Array at $\sim1$\arcsec\, resolution. The PV diagrams are reploted in Figure \ref{fig:i04166_obs_co32}, which show finer structures especially in the blue-shifted lobe.
    
    The velocity feature between $\sim1500$--$2000$ AU in the blue-shifted lobe is particularly interesting. Unlike the main tooth pattern that are likely associated with a spherical wind-like geometry, the emission between them appears to show a reverse pattern. It may be possible that it represents the true velocity variation over time. In this case, the increase of velocity with distance may be interpreted as a decrease of outflow launching velocity with time over the particular era. Unfortunately, the emission is weak and fragmented and would need to be confirmed by higher sensitivity observation before further modeling could help understand its nature. The innermost red-shifted emission is also interesting. There seems to be multiple overlapping teeth pattern within 1000 AU from the source and it is possible that we are looking at a series of ejecta launched with increasingly higher speed. Still, higher sensitivity data is needed to confirm its behavior.     
	
	The fact that the outer knots do not preserve clues of the detailed velocity variation is actually a benefit when testing whether the pattern is consistent with a spherical wind scenario or not. After gas of different velocities merges into a single packet and move more-or-less at a constant speed, the observed sawtooth patterns will most likely be associated with lateral velocity components and we can worry less about the velocity variation along the flow direction. The model predictions are hence more reliable for the other knots. Establishing the wind nature of the velocity field can also help interpret the detailed velocity structures found close to the source. Without such knowledge, it would be difficult to tell whether the tooth-like patterns in the inner knots in Figure \ref{fig:i04166_obs_co32} are signs of a wind-like velocity field or an axial velocity variation. In this sense, the sawtooth velocity pattern of the I04166 jet poses an unique clue for the study of its ejection history.  

    Similar quasi-linear velocity patterns have also been seen in some other Class 0 outflows. SMA \sio{8}{7} and \co{3}{2} observations toward the L1448C outflow revealed tooth-like velocity patterns in at least two pairs of its EHV peaks \citep[see Figure 7 of][]{hirano2010}. These ``fast tail--slow head'' features might have the same geometric origin as the I04166 outflow, but it would be difficult to confirm. Unlike I04166, we do not have multiple knots distributed across a large distance. Several quasi-linear features have also been identified in the HH 211 molecular jet \citep[see Figure 7 of][]{lee2009}. In particular, the \sio{8}{7} emission of the blue-shifted knots BK2 and BK3 present a ``fast tail--slow head'' velocity structure that were interpreted as the intrinsic velocity variation along the jet axis during a catching-up process \citep{jhan2016}. It would be interesting to see whether, or how much, perpendicular velocity component (either due to geometry or from sideway ejection of an internal working surface) contributes to the observed velocity pattern. The interpretation may be further complicated by the wiggling of HH 211 jet \citep{moraghan2016}.  
    In any case, disentangling contributions from different potential mechanisms is critical for a thorough understanding, and would require high quality observations toward many different outflows. 
    
\section{CONCLUSION}	
	
	\label{sec:conclusion}
	
	We propose that the observed sawtooth-like velocity pattern of I04166 molecular jet can be associated with an expanding spherical wind geometry of density concentrating near the axis, similar to that predicted by the X-wind model. We derive an analytical formula for the pattern slope, which describes its dependence on the outflow inclination angle, the ejecta velocity, and its distance from the driving center. The formula can reasonably explain the systematic variation of slopes across the whole pattern using one same underlying outflow inclination angle, which is not automatically explained in the paradigm of jet and internal working surface.  The reasonable agreement is therefore a quite compelling signpost of an underlying spherical wind-like velocity field in the I04166 jet. The dimensionless slope of the tooth-like velocity pattern can also be used to estimate the outflow inclination in this picture. 
	
	Using numerical simulations of variable velocity wind in framework of the unified wind model of \citet{shang2006}, we are able to generate synthetic \co{2}{1} PV diagrams that reproduce the observed pattern well. This realistically demonstrates the feasibility of the proposed wind scenario. 
	While a spherical ejecta picture can nicely describe the kinematics of the outer knots, clues of the velocity history of general interest are better preserved in the inner knots where gas of different speeds has not yet completely merge together. Different potential mechanisms may underlie the observed velocity pattern, including the intrinsic velocity variation along the flow direction, the potential side-way velocity from internal working surfaces of shocks, and the diverging (wind-like) geometry of the flow. A correct model of flow geometry is therefore critical for an accurate understanding of the flow history. The unique sawtooth pattern of the I04166 jet has allowed us to associate it with a wind interpretation, which could help disentangle the velocity pattern in the innermost region. High resolution high sensitivity observations could help confirm the detailed flow structure and shed light on its velocity history.

\acknowledgments

We thank M. Tafalla and J. Santiago-Garc\'ia for providing the \co{2}{1} data of the molecular outflows IRAS 04166$+$2706. 
We also thank an anonymous referee whose comments have helped improve the clarity and presentation of the manuscript.
This paper makes use of the following ALMA data: ADS/JAO.ALMA\#2012.1.00304.S. ALMA is a partnership of ESO (representing its member states), NSF (USA) and NINS (Japan), together with NRC (Canada), MOST and ASIAA (Taiwan), and KASI (Republic of Korea), in cooperation with the Republic of Chile. The Joint ALMA Observatory is operated by ESO, AUI/NRAO and NAOJ. 
This work also uses data from The Submillimeter Array. The Submillimeter Array is a joint project between the Smithsonian Astrophysical Observatory and the Academia Sinica Institute of Astronomy and Astrophysics and is funded by the Smithsonian Institution and the Academia Sinica.
This work utilized tools developed by the CHARMS group and computing resources and cluster in TIARA at Institute of Astronomy and Astrophysics in Academia Sinica (ASIAA). We also acknowledge grant support from Ministry of Science and Technology in Taiwan by MOST 102-2119-M-001-008-MY3 and MOST 105-2119-M-001-037-.


\bibliographystyle{aasjournal}
\bibliography{Outflows}

\begin{thebibliography}{}
\expandafter\ifx\csname natexlab\endcsname\relax\def\natexlab#1{#1}\fi

\bibitem[{{Andre} {et~al.}(1993){Andre}, {Ward-Thompson}, \&
  {Barsony}}]{andre1993}
{Andre}, P., {Ward-Thompson}, D., \& {Barsony}, M. 1993, \apj, 406, 122

\bibitem[{{Bachiller}(1996)}]{bachiller1996}
{Bachiller}, R. 1996, \araa, 34, 111

\bibitem[{{Bally}(2016)}]{bally2016}
{Bally}, J. 2016, \araa, 54, 491

\bibitem[{{Frank} {et~al.}(2014){Frank}, {Ray}, {Cabrit}, {Hartigan}, {Arce},
  {Bacciotti}, {Bally}, {Benisty}, {Eisl{\"o}ffel}, {G{\"u}del}, {Lebedev},
  {Nisini}, \& {Raga}}]{frank2014}
{Frank}, A., {Ray}, T.~P., {Cabrit}, S., {et~al.} 2014, Protostars and Planets
  VI, 451

\bibitem[{{Gueth} \& {Guilloteau}(1999)}]{gueth1999}
{Gueth}, F., \& {Guilloteau}, S. 1999, \aap, 343, 571

\bibitem[{{Haro}(1952)}]{haro1952}
{Haro}, G. 1952, \apj, 115, 572

\bibitem[{Hartmann {et~al.}(2016)Hartmann, Herczeg, \& Calvet}]{hartmann2016}
Hartmann, L., Herczeg, G., \& Calvet, N. 2016, \araa, 54, 135

\bibitem[{{Herbig}(1951)}]{herbig1951}
{Herbig}, G.~H. 1951, \apj, 113, 697

\bibitem[{{Hirano} {et~al.}(2010){Hirano}, {Ho}, {Liu}, {Shang}, {Lee}, \&
  {Bourke}}]{hirano2010}
{Hirano}, N., {Ho}, P.~P.~T., {Liu}, S.-Y., {et~al.} 2010, \apj, 717, 58

\bibitem[{{Ho} {et~al.}(2004){Ho}, {Moran}, \& {Lo}}]{ho2004}
{Ho}, P.~T.~P., {Moran}, J.~M., \& {Lo}, K.~Y. 2004, \apjl, 616, L1

\bibitem[{Hodapp {et~al.}(2012)Hodapp, Chini, Watermann, \& Lemke}]{hodapp2012}
Hodapp, K.~W., Chini, R., Watermann, R., \& Lemke, R. 2012, \apj, 744, 56

\bibitem[{Jhan \& Lee(2016)}]{jhan2016}
Jhan, K.-S., \& Lee, C.-F. 2016, \apj, 816, 32

\bibitem[{{Lee} {et~al.}(2009){Lee}, {Hirano}, {Palau}, {Ho}, {Bourke},
  {Zhang}, \& {Shang}}]{lee2009}
{Lee}, C.-F., {Hirano}, N., {Palau}, A., {et~al.} 2009, \apj, 699, 1584

\bibitem[{{Lee} {et~al.}(2007){Lee}, {Ho}, {Hirano}, {Beuther}, {Bourke},
  {Shang}, \& {Zhang}}]{lee2007}
{Lee}, C.-F., {Ho}, P.~T.~P., {Hirano}, N., {et~al.} 2007, \apj, 659, 499

\bibitem[{{Li} \& {Shu}(1996)}]{li1996}
{Li}, Z.-Y., \& {Shu}, F.~H. 1996, \apj, 472, 211

\bibitem[{Moraghan {et~al.}(2016)Moraghan, Lee, Huang, \&
  Vaidya}]{moraghan2016}
Moraghan, A., Lee, C.-F., Huang, P.-S., \& Vaidya, B. 2016, \mnras, 460, 1829

\bibitem[{{Raga} {et~al.}(1990){Raga}, {Binette}, {Canto}, \&
  {Calvet}}]{raga1990b}
{Raga}, A.~C., {Binette}, L., {Canto}, J., \& {Calvet}, N. 1990, \apj, 364, 601

\bibitem[{{Santiago-Garc{\'{\i}}a} {et~al.}(2009){Santiago-Garc{\'{\i}}a},
  {Tafalla}, {Johnstone}, \& {Bachiller}}]{santiago-garcia2009}
{Santiago-Garc{\'{\i}}a}, J., {Tafalla}, M., {Johnstone}, D., \& {Bachiller},
  R. 2009, \aap, 495, 169

\bibitem[{{Shang} {et~al.}(2006){Shang}, {Allen}, {Li}, {Liu}, {Chou}, \&
  {Anderson}}]{shang2006}
{Shang}, H., {Allen}, A., {Li}, Z.-Y., {et~al.} 2006, \apj, 649, 845

\bibitem[{{Shu} {et~al.}(1995){Shu}, {Najita}, {Ostriker}, \&
  {Shang}}]{shu1995}
{Shu}, F.~H., {Najita}, J., {Ostriker}, E.~C., \& {Shang}, H. 1995, \apjl, 455,
  L155

\bibitem[{{Shu} {et~al.}(2000){Shu}, {Najita}, {Shang}, \& {Li}}]{shu2000}
{Shu}, F.~H., {Najita}, J.~R., {Shang}, H., \& {Li}, Z.-Y. 2000, Protostars and
  Planets IV, 789

\bibitem[{{Stone} \& {Norman}(1993)}]{stone1993b}
{Stone}, J.~M., \& {Norman}, M.~L. 1993, \apj, 413, 210

\bibitem[{{Tafalla} {et~al.}(2004){Tafalla}, {Santiago}, {Johnstone}, \&
  {Bachiller}}]{tafalla2004}
{Tafalla}, M., {Santiago}, J., {Johnstone}, D., \& {Bachiller}, R. 2004, \aap,
  423, L21

\bibitem[{{Tafalla} {et~al.}(2017){Tafalla}, {Su}, {Shang}, {Johnstone},
  {Zhang}, {Santiago-Garc{\'{\i}}a}, {Lee}, {Hirano}, \& {Wang}}]{tafalla2017}
{Tafalla}, M., {Su}, Y.-N., {Shang}, H., {et~al.} 2017, \aap, 597, A119

\bibitem[{{Wang} {et~al.}(2015){Wang}, {Shang}, {Krasnopolsky}, \&
  {Chiang}}]{wang2015}
{Wang}, L.-Y., {Shang}, H., {Krasnopolsky}, R., \& {Chiang}, T.-Y. 2015, \apj,
  815, 39

\bibitem[{{Wang} {et~al.}(2014){Wang}, {Shang}, {Su}, {Santiago-Garc{\'{\i}}a},
  {Tafalla}, {Zhang}, {Hirano}, \& {Lee}}]{wang2014}
{Wang}, L.-Y., {Shang}, H., {Su}, Y.-N., {et~al.} 2014, \apj, 780, 49

\bibitem[{{Wilson}(1984)}]{wilson1984}
{Wilson}, M.~J. 1984, \mnras, 209, 923

\bibitem[{Yoo {et~al.}(2017)Yoo, Lee, Mairs, Johnstone, Herczeg, Kang, Kang,
  Cho, \& Team}]{yoo2017}
Yoo, H., Lee, J.-E., Mairs, S., {et~al.} 2017, \apj, 849, 69

\end{thebibliography}

\end{document}